\documentclass [conference]{IEEEtran}
\usepackage{algorithm,amsbsy,amsmath,amssymb,epsfig,bbm,mathrsfs,multirow,amsthm}
\usepackage{color}
\usepackage{algpseudocode}
\usepackage{graphicx,caption,subcaption,array,multirow,url}
\usepackage{makecell,cite}
\DeclareMathOperator*{\argmax}{arg\,max}

\DeclareMathAlphabet{\mathpzc}{OT1}{pzc}{m}{it}

\newtheorem{theorem}{\textbf{\textsc{Theorem}}}

\usepackage[top=0.71in, bottom=1in, left=0.65in, right=0.65in]{geometry}
\begin{document}
	\title{MetaChain: A Novel Blockchain-based Framework for Metaverse Applications}
	\author{Cong T. Nguyen, Dinh Thai Hoang, Diep N. Nguyen and Eryk Dutkiewicz  \\		
		 School of Electrical and Data Engineering, University of Technology Sydney, Australia \\
		\vspace{-5mm}	}	
	
\maketitle
\thispagestyle{empty}
	\begin{abstract}	

	Metaverse has recently attracted paramount attention due to its potential for future Internet. However, to fully realize such potential, Metaverse applications have to overcome various challenges such as massive resource demands, interoperability among applications, and security and privacy concerns. In this paper, we propose MetaChain, a novel blockchain-based framework to address emerging challenges for the development of Metaverse applications. In particular, by utilizing the smart contract mechanism, MetaChain can effectively manage and automate complex interactions among the Metaverse Service Provider (MSP) and the Metaverse users (MUs). In addition, to allow the MSP to efficiently allocate its resources for Metaverse applications and MUs' demands, we design a novel sharding scheme to improve the underlying blockchain's scalability. Moreover, to leverage MUs' resources as well as to attract more MUs to support Metaverse operations, we develop an incentive mechanism using the Stackelberg game theory that rewards MUs' contributions to the Metaverse. Through numerical experiments, we clearly show the impacts of the MUs' behaviors and how the incentive mechanism can attract more MUs and resources to the Metaverse.
	

	\end{abstract}
	{\it Keywords-} Blockchain, Metaverse, sharding, and Stackelberg game.
	
	\thispagestyle{empty}
	\section{Introduction}

	 The term Metaverse refers to next-generation Internet applications that aim to create virtual 3D environments where humans can interact with each other as well as the applications' functionalities via digital avatars~\cite{meta2}. Although the original concept dates back to 1992~\cite{meta2}, Metaverse has recently attracted paramount attention due to a huge potential for future Internet.
	Facebook has recently announced a \$10 billion investment for Metaverse development, and at the same time, Microsoft has developed a Metaverse meeting framework~\cite{web}. Roblox, a \$38 billion company, has recently declared its goal to build a single Metaverse framework for millions of users~\cite{web}. All these examples foretell a future where Metaverse applications will rival, or even replace, conventional Internet applications. However, to realize the potentials and benefits of such massive virtual environments, Metaverse applications have to overcome various challenges such as massive resource demands, ultra-low latency requirements, interoperability among applications, and security and privacy concerns.

	To address the aforementioned challenges, blockchain has been considered to be a promising solution~\cite{meta2}. In particular, thanks to the smart contract mechanisms~\cite{SC}, blockchain can manage and automate complex interactions among various entities in Metaverse, such as Metaverse Service Providers (MSPs), users, and digital content creators. Moreover, with outstanding benefits of immutability and transparency~\cite{PoS}, blockchain can play a key role in ensuring data integrity and protecting digital assets in Metaverse applications. Furthermore, with the asymmetric key and digital signature mechanisms~\cite{PoS}, blockchain can enhance the users' privacy and anonymity~\cite{meta2}. However, blockchain, especially when applied in Metaverse, also faces various challenges~\cite{meta2}. Particularly, most of the current blockchain networks have been employing the Proof-of-Work (PoW)~\cite{Bitcoin} consensus mechanism which has a significant delay, e.g., Bitcoin requires roughly one hour to confirm a transaction~\cite{PoS}. Moreover, scalability is another challenge of the PoW mechanism. Particularly, due to the requirement of the PoW mechanism, it is difficult to expand the blockchain network to improve its transaction processing capability. Consequently, these limitations are the major obstacles to the future implementation of blockchain-based Metaverse applications that are expected to provide time-sensitive services to millions of users.
	
	Since Metaverse has just recently attracted attention, research on this topic is still quite limited. Most of the existing works focus on resource allocations for Metaverse, such as~\cite{re2,re3,re4,re5}. Although a few proposed frameworks suggest using blockchain, e.g.,~\cite{re2,re3}, none of them attempts to utilize blockchain to facilitate Metaverse interactions, e.g., digital asset tradings and service selling. Moreover, blockchain's scalability issue is not addressed in both~\cite{re2} and~\cite{re3}. If this issue still persists, blockchain might become the bottleneck in future Metaverse applications that involve millions or even billions of users.

			\begin{figure}[!]
		\includegraphics[width=.5\textwidth]{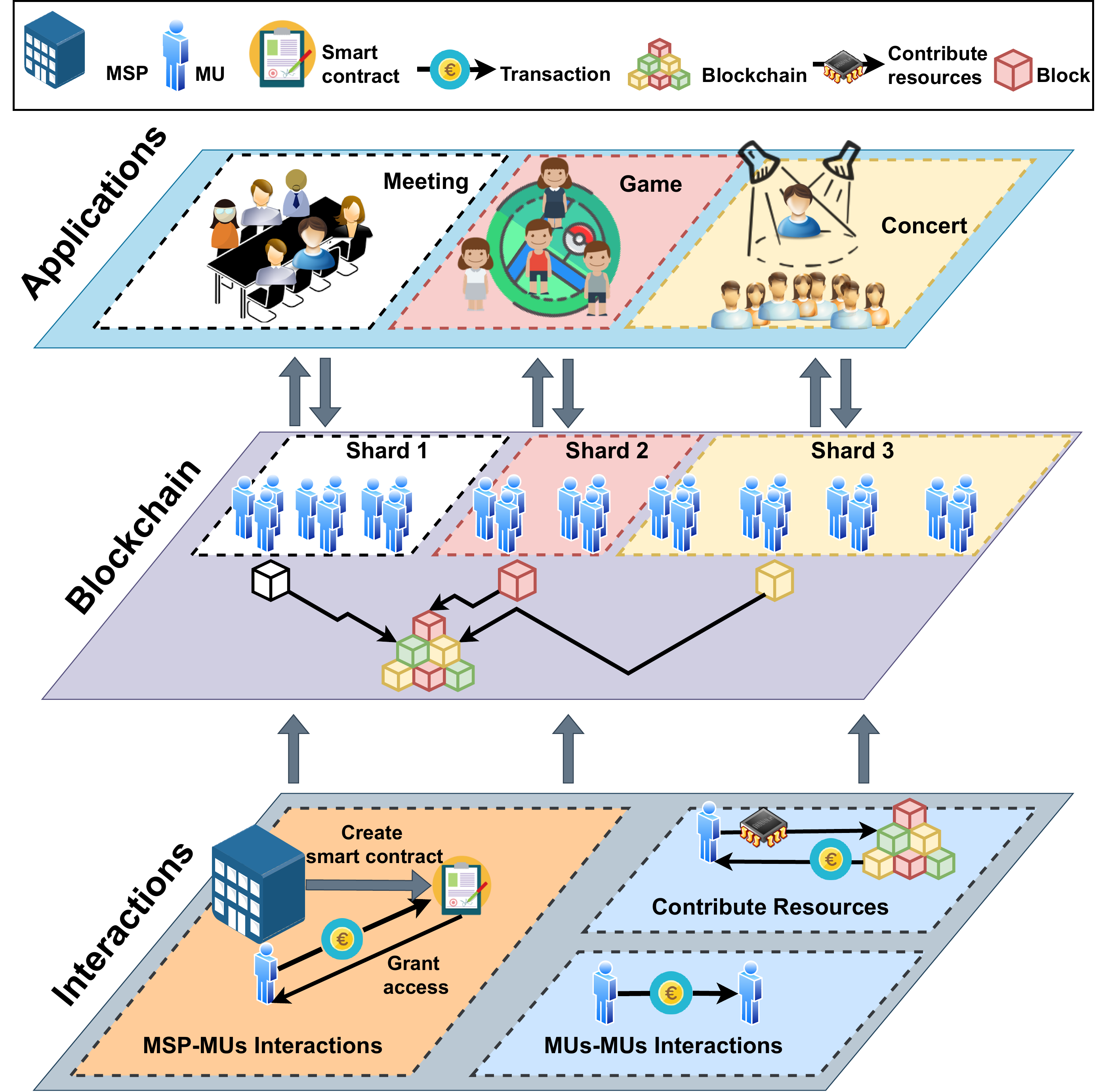}
		\centering
		\caption{An illustration of the proposed system.}
		\label{Fig:framework}
	\end{figure}

	To address the abovementioned challenges, we develop MetaChain, a novel blockchain-based framework that can (i) facilitate various interactions between the MSP and the Metaverse users (MUs), (ii) efficiently manage resources according to application demands, and (iii) encourage MUs to contribute resources to the blockchain and Metaverse applications, bringing benefits to both the MUs and the MSP. In particular, MetaChain utilizes smart contract mechanisms to enable and automate various interactions among the MSP and the MUs without requiring a trusted third party. Moreover, to improve the blockchain's scalability, we propose a sharding scheme~\cite{shard} in which the shard numbers and sizes are decided based on the Metaverse applications' demands, thereby allowing the MSP to allocate resources dynamically. Furthermore, since the Metaverse applications often require significant computational resources, we design an incentive mechanism using the Stackelberg game theory~\cite{Game}. With this mechanism, the MSP can leverage MUs' resources to support the blockchain operations and the Metaverse applications. At the same time, the MUs can earn rewards for their contribution, and thus the incentive mechanism can also attract more MUs to the Metaverse. Through numerical experiments, we clearly show the impacts of the MUs' behaviors and how the incentive mechanism can attract more MUs and resources to the Metaverse.
	
	\section{MetaChain}
	\label{sec:SM}
\subsection{System Overview}
Fig.~\ref{Fig:framework} demonstrates an overview of MetaChain. In this framework, the MSP is a service provider that offers their applications/services, e.g., VR games, virtual conferences, and virtual concerts, to the MUs. Each MU has an account and corresponding assets, e.g., tokens, coins, and items, in the system. The MUs can pay the MSP to gain access to the service as well as transfer digital assets to other MUs. Furthermore, the MUs can contribute computational resources to the blockchain to earn rewards. The blockchain, managed by the MSP, serves as a platform to manage all interactions between the MSP and the MUs as well as interactions among the MUs. Those interactions, along with the MUs' assets, are recorded in the blockchain in terms of transactions.
%
%

\subsection{MSP-to-MUs Interactions}
Most of the interactions between the MSP and the MUs can be done via the blockchain. This helps to improve the MUs' privacy and anonymity, as well as automating processes. For example, the MSP can broadcast its fees and options (e.g., number of people and themes) for Metaverse meeting services on the blockchain in terms of a smart contract, i.e., a user-defined program which is automatically enforced when predefined conditions are met~\cite{SC}. Then, an MU can send a transaction that specifies various options (e.g., participants and time) to the smart contract. If the payment is sufficient, the smart contract is triggered to automatically send confirmation transactions to the involved participants. These transactions can later be used as a proof to gain access to the virtual meeting room.

\subsection{MUs-to-MUs Interactions}
Besides automating processes, the blockchain can also serve as an immutable database to record the MUs' transfers of assets. Moreover, the framework also supports transactions made among the MUs without the need for a trusted authority. Via the blockchain, digital assets can be transferred and exchanged smoothly among the MUs. For example, similar to real-life scenarios, an MU who purchases a virtual concert ticket but cannot attend can resell it to another MU. The proposed framework can support such exchanges in a secure and transparent manner via smart contracts or blockchain transactions.

\subsection{Blockchain Shardings}
Like many other social networks (e.g., Facebook and Twitter), Metaverse applications are expected to provide services for millions of MUs, which is very challenging for conventional PoW blockchain networks. Therefore, we propose to implement the sharding mechanism~\cite{shard}. The sharding mechanism divides the blockchain network into smaller sub-networks. Each sub-network has its own consensus process to create blocks to add to the chain. Since there are multiple shards in the network, blocks can be created multiple times faster than a network without shards. As the application grows in demand, more shards can be created to ensure a high Quality-of-Service to MUs.

Unlike the conventional sharding mechanism where it is only employed to increase the transaction throughput~\cite{shard}, we propose to create shards based on the actual demands in terms of both transaction processing capability and computational resource demands. Particularly, in applications such as virtual conferences, the MUs' demand may fluctuate significantly depending on the time and geographical location, e.g., when most conferences commence in Europe, most MUs in East Asia are sleeping. Moreover, the MSP may offer multiple services, but some of them may have much more MUs than the others. Therefore, we propose to create shards according to the MUs' demands, e.g., one shard per region/application. In this way, the MSP can dynamically allocate their resources properly to each shard according to the demands.
\subsection{MU Resource Contribution}
Furthermore, Metaverse applications are often characterized by their huge computational resources demands, e.g., to create virtual environments. Although the rapid development of user devices' capacity can partially circumvent this issue, there are still applications where the MUs need a unified environment to interact with each other, e.g., gaming. Therefore, we propose to utilize the blockchain shards to create a framework for the MUs to contribute their computational resources to the Metaverse applications in exchange for digital assets, e.g., tokens or service access. With this approach, the MSP can leverage the MUs' resources and also attract more MUs to the network. In particular, depending on each shard's demand, the MSP can set a payment rate for the shard. MUs who contribute resources to the shards can be paid in the form of network tokens according to their contributions, e.g., using the pay-per-share protocol~\cite{pps}.

Nevertheless, attracting and incentivizing the MUs to contribute resources to the shards is a challenging task. The reason is that the MSP cannot directly allocate the MU resources to the shards. Instead, the MSP can only incentivize the MUs using the payment rate. Moreover, since the MUs are rational (i.e., they aim to maximize their profits), they may concentrate their resources on the shards with higher payment rates. Furthermore, with the use of the pay-per-share protocol, MUs may have conflicts-of-interests, i.e., the more MUs join a shard, the less reward each MU gets. Therefore, we will examine the MUs and the MSP strategy using game theory in the next Section to show how such rational behaviors affect the proposed system and how the MSP can set the payment rate to properly allocate MU resources. 
	\section{Stackelberg Game Analysis}
	\subsection{MUs and Shards}
	The proposed system consists of a blockchain divided into a set $\mathcal{M}$ of $M$ shards according to the number of Metaverse applications. There is a set $\mathcal{N}$ of $N$ MUs who want to contribute resources to the blockchain and Metaverse applications. The MUs have computational resource capacities $\mathbf{R} = (R_1, \ldots, R_N)$. In the proposed system, each MU decides how much resources, denoted by $r_n^m$, it wants to contribute to each shard. Correspondingly, there is a unit operational cost at each MU when it contributes resources, denoted by $C_n$. The MSP can decide the total amount of payment for the shards, denoted by $\mathbf{P} = (P_1, \ldots, P_M)$. This payment can be paid to the MUs in the forms of network tokens (coins) via the block reward mechanism~\cite{PoS} or direct transaction from the MSP. 

In practice, the MSP often announces the payment for each shard first. Based on that, the MUs will decide how
much resource to contribute to each shard. Therefore, the interaction between	the MSP and the MUs can be formulated as a single-leader multi-followers Stackelberg game model~\cite{Game}, denoted by $\mathcal{G}$. In $\mathcal{G}$, there is one leader (the MSP) who first declares its strategy, i.e., $P_m$, and then the followers (MUs) will decide how much resources to contribute. Let $\mathcal{S}_n$ and $s^*_n$ denote the set of all possible strategies of follower $n$ and the follower's best response, respectively. $s^*_n$ is the strategy that yields the best payoff given a fixed strategy $s_L$ of the leader, i.e.,
	\begin{equation}
		\label{eq:bestresponse}
		U_n(s^*_n,s_L)\geq U_i(s'_n,s_p), \forall s'_n \in \mathcal{S}_n,
	\end{equation}
	where $U_n(s^*_n,s_L)$ is the utility (payoff) function of follower $n$. Based on the follower's best response, the Stackelberg strategy for the leader is a strategy $s^*_L$ such that
	\begin{equation}
		\label{eq:leader}
		s^*_L= \argmax_{s_L=\textbf{P}} U_L(s_L,s^*_n),
	\end{equation}
	where $U_L(s_L,s^*_n)$ is the utility function of the leader. The Stackelberg solution then can be defined by a tuple $(s^*_L,s^*_n)$, and its corresponding utility tuple $(U^*_p, U^*_i)$ is the Stackelberg equilibrium of $\mathcal{G}$. $\mathcal{G}$ can be divided into two stages (sub-games). At the first stage, denoted by $\mathcal{G}_L$, the leader announces its strategy. Then, at the second stage, denoted by $\mathcal{G}_F$, the followers determine their strategies in response to the leader's strategy. In the following, we use backward-induction-based analysis~\cite{Game} to determine the Stackelberg equilibrium of this game.
	\subsection{Stage II - Followers' Sub-game $\mathcal{G}_F$}
	In the considered system, MU $n$ can freely contribute its resources $r_n^m$ to every shard $m, \forall m \in \mathcal{M}$, such that $\sum_{m}^{M} r^m_n \leq R_n$. Each shard has a fixed amount of payment which will be fairly distributed among the MUs based on their proportions of contributed resources (e.g., using the pay-per-share protocol~\cite{pps}). The follower's payoff from contributing $r_n^m$ to shard $m$ is:
	\begin{equation}
		\label{eq:payoffpool}
		U^m_n=\dfrac{r_n^m}{r^m_n+\sum_{i \in \mathcal{N}_{-n}} r^m_i}P_m,
	\end{equation}
where $\mathcal{N}_{-n}$ is the set of all followers except follower $n$. Then, the follower's total utility can be determined by
	\begin{equation}
	\label{eq:payoff}
	U_n=\sum_{m=1}^M U^m_n=\sum_{m=1}^M \dfrac{r_n^m}{r^m_n+\sum_{i \in \mathcal{N}_{-n}} r^m_i}P_m -C_n\sum_{m=1}^M r_n^m.
	\end{equation}
	We first examine the existence and uniqueness of the $\mathcal{G}_F$'s equilibrium in Theorem 1. 
	\begin{theorem}
		There exists a unique equilibrium in $\mathcal{G}_F$.
	\end{theorem}
	\begin{IEEEproof}
The proof is provided in Appendix A.
	\end{IEEEproof}
	\subsection{Stage I - Leader's sub-game $\mathcal{G}_L$}
	For the MSP, we propose the utility function as follow
	\begin{equation}
U_L=\sum_{m=1}^{M} \bigg(\alpha_m \ln\bigg(\sum_{n=1}^{N} r_n^m\bigg) -P_m\bigg),
	\end{equation}
where $\alpha_m$ is a control parameter that represents the priority of the MSP in terms of resource allocation for each shard, e.g., a shard that has higher resources demand has a higher $\alpha$. Moreover, the log function is used to prevent too much resources concentrated into a single shard.

Then, the optimal strategy $s^*_L$ of the leader, which is the set of payments that yields the highest payoff given the best responses of all	followers, can be defined by
		\begin{equation}
		\label{eq:leader2}
		\begin{split}
		s^*_L&= \argmax_{s_L=\textbf{P}} U_L(s_L,s^*_n)=\sum_{m=1}^{M} \bigg(\alpha_m \sum_{n=1}^{N} r_n^{*m} -P_m\bigg),\\
		\end{split}
	\end{equation}
To find the Stackelberg equilibrium for $\mathcal{G}$, we develop an iterative algorithm as described in Algorithm 1, which works as follows. At the first iteration, parameters and variables are initialized. Then, a function $\Call{CALCULATE $U_L$}{\textbf{P},max}$ is called to find the utility of the leader given the current $\textbf{P}$. Then, iteratively each shard reward is increased by 1 and $\Call{CALCULATE $U_L$}{\textbf{P},max}$ is called again to find $U_L$. When a new optimal value for $U_L$ is found, the value and the corresponding strategy are recorded. The algorithm is stopped when the stopping criterion is met. To find the best response of the followers in $\Call{CALCULATE $U_L$}{\textbf{P},max}$, we first fix the strategy of all followers except follower $n$. Then, we use the function \textit{fmincon} in Matlab to find the optimal strategy $\textbf{r}^*_n$ of follower $n$. Next, we fix the strategy of follower $n$ and continue to find the optimal strategy of follower $n+1$. This loop is stopped when no follower changes their strategy.
	\begin{algorithm}
	\caption{Iterative Algorithm to find the Stackelberg equilibrium}\label{euclid}
	\begin{algorithmic}[1]
		\Repeat
		\State $max\gets 0$, $P_m \gets 1$,$\forall m \in \mathcal{M}$, $\textbf{P}^* \gets \textbf{P}$
		
		\State \Call{Calculate $U_L$}{\textbf{P},$max$}
		\For {$i:=1$ to M}
		\State $P_i\gets P_i+1$
		\State \Call{Calculate $U_L$}{\textbf{P},$max$}
		\EndFor
		\Until {Stopping criteria}
				\Function{Calculate $U_L$}{\textbf{P},$max$}
		\Repeat
		\For {$n:=1$ to $N$}
		\State Find $\textbf{r}^*_n$ 
		\EndFor 
		\Until{No follower changes strategy}
		\If{$U_L>max$}     
		\State $max \gets U_L$, $\textbf{P}^* \gets \textbf{P}$
		\EndIf  
		\State \Return $max$
		\EndFunction
	\end{algorithmic}
\end{algorithm}

	\section{Numerical Results}
	\label{sec:simu}
	In this Section, we simulate a system with 4 MUs (followers) and a blockchain with 2 shards to evaluate their strategies and their impacts on the MSP's strategy and utility. The resource capacity of each MU is set as $\mathbf{R}=[100,200,300,500]$ and their unit costs are $\textbf{C}=[0.2,0.1,0.3,0.2]$. The priority of each shard is set at $\boldsymbol{\alpha}=[4,6]$. Under this parameter setting, we first examine the utility function of one MU and show the MU's optimal strategy when the other MUs' and the MSP's strategies are fixed. Then, we use the function $\Call{CALCULATE}{\textbf{P},max}$ to show how the MUs' strategies converge to the unique $\mathcal{G}_F$'s equilibrium. Finally, we simulate the game using Algorithm 1 to illustrate the utility function of the MSP and the Stackelberg equilibrium.  
	
\begin{figure}[!t]
			\includegraphics[width=.45\textwidth]{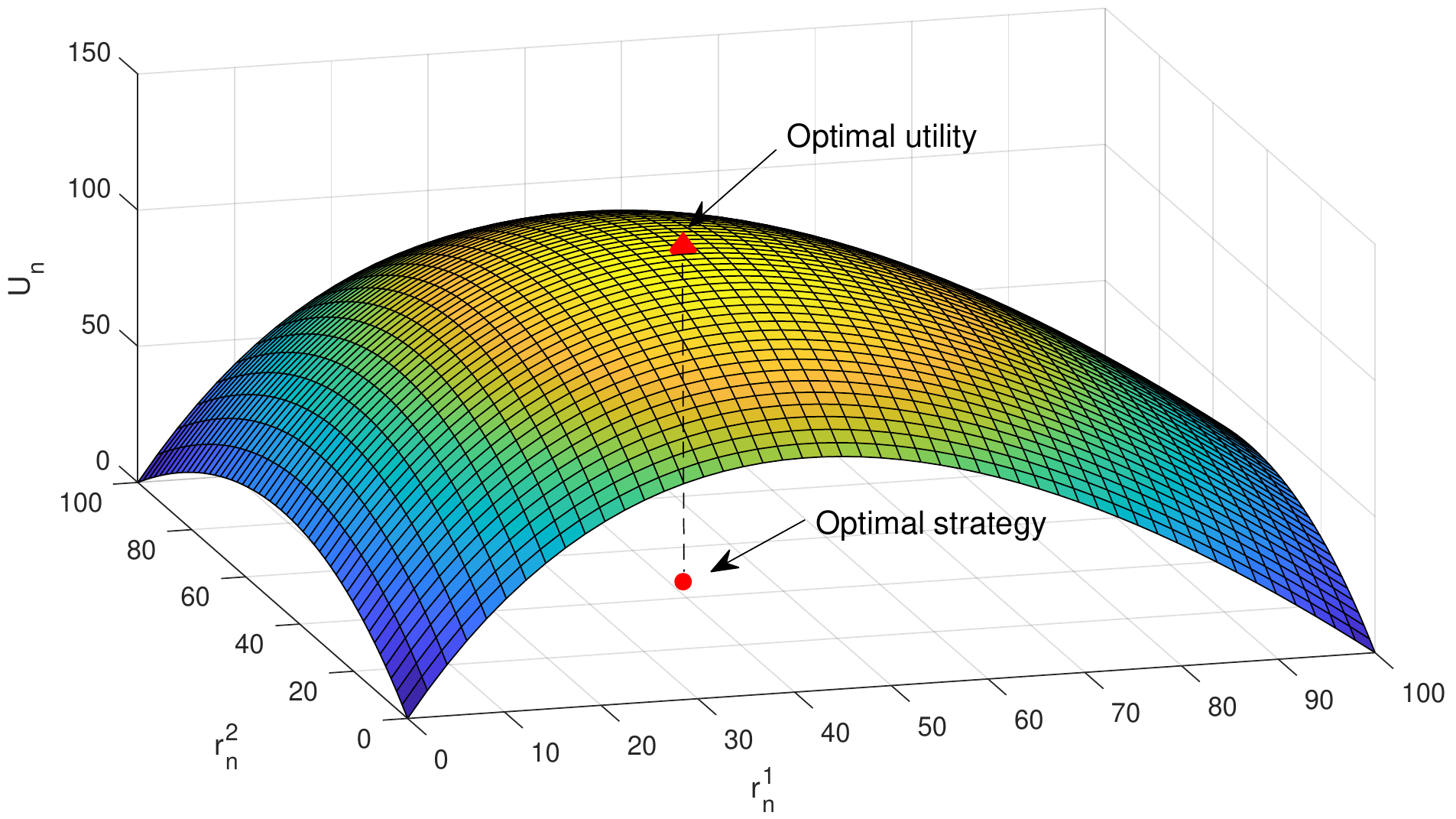}
		\centering
		\caption{An illustration of the MU's utility function.}
		\label{Fig:utif}
	\end{figure}

	Fig.~\ref{Fig:utif} illustrates the utility function and the optimal strategy of MU $n$ when the strategies of the other MUs and the MSP are fixed. In this example, we fix the payment as $\textbf{P}=[1000,2000]$, the strategies of other MUs as $\textbf{T}=[100,300]$, and the unit cost of MU $n$ as $C_n=5$. As observed from the figure, the MU can achieve an optimal utility $U^*_n=121.68$ when it contributes $r_n^1=41.42$ to shard 1 and $r_n^2=46.41$ to shard 2. It is worth noting that although we set the MU's resource capacity to $R_n=100$, the MU does not contribute all of its resources to the shards. This is because after a certain threshold, contributing more resources brings less payoff than the cost it incurs. This threshold depends on MU $n$'s unit cost as well as the strategies of the other MUs and the MSP.

\begin{figure}[!t]
	 	\includegraphics[width=.45\textwidth]{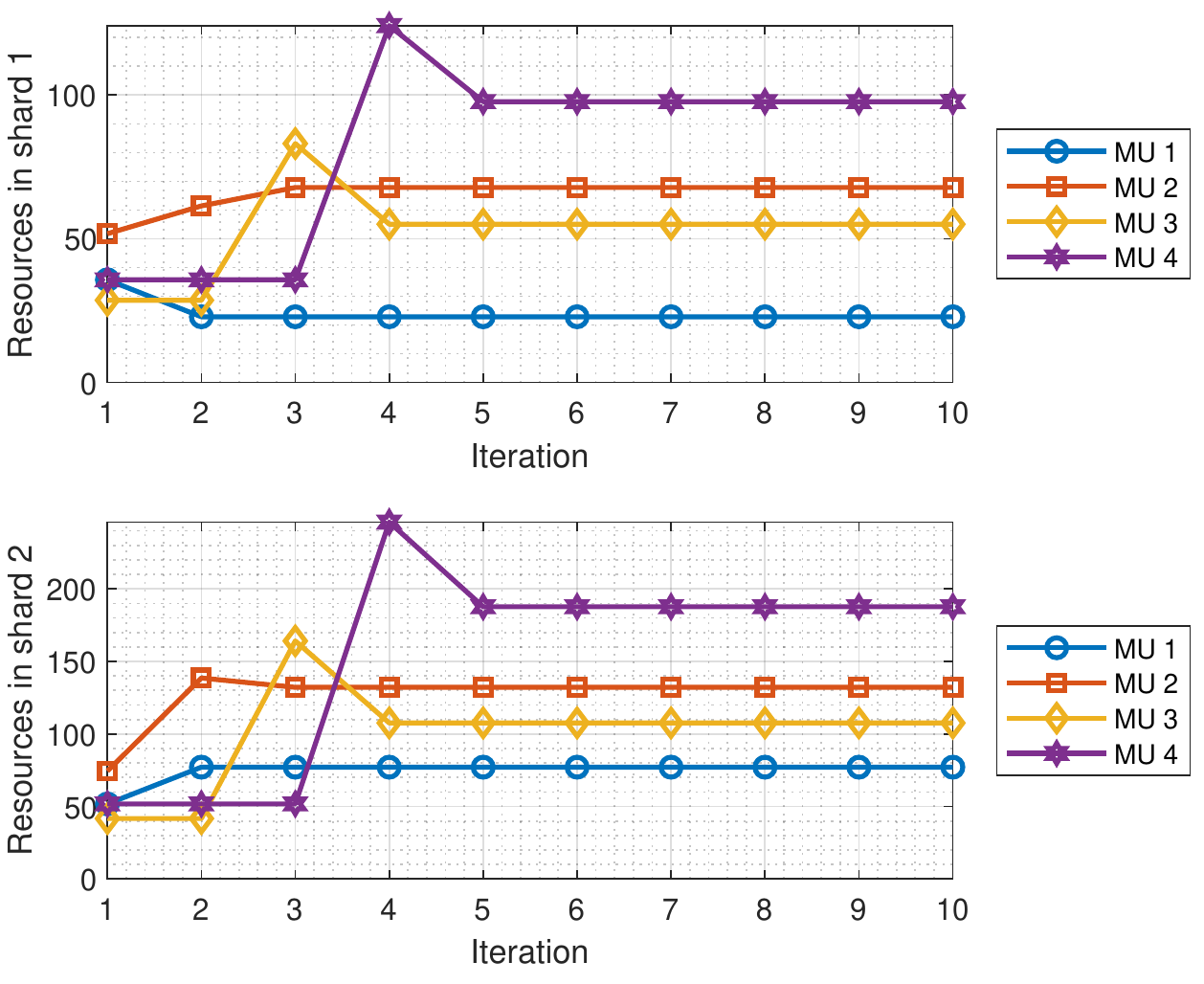}
	 	\centering
	 	\caption{MUs' strategies converge to the equilibrium.}
	 	\label{Fig:conv}
	 \end{figure}

	Fig.~\ref{Fig:conv} illustrates how the MUs' strategies converge to the equilibrium. In this example, we fix the shard's payment as $\textbf{P}=[100,200]$. As observed from the figure, the MUs' strategies converge after 5 iterations. At the equilibrium, we can observe that MUs 1 and 2 contribute all their resources to the shards, while MUs 3 and 4 do not. The reason is MUs 1 and 2 have lower unit costs, and thus they can gain more benefits per unit of resources. Moreover, due to the difference in unit cost, MU 2 contributes more resources even though it has a lower capacity than that of MU 3.

\begin{figure}[!]
		\includegraphics[width=.4\textwidth]{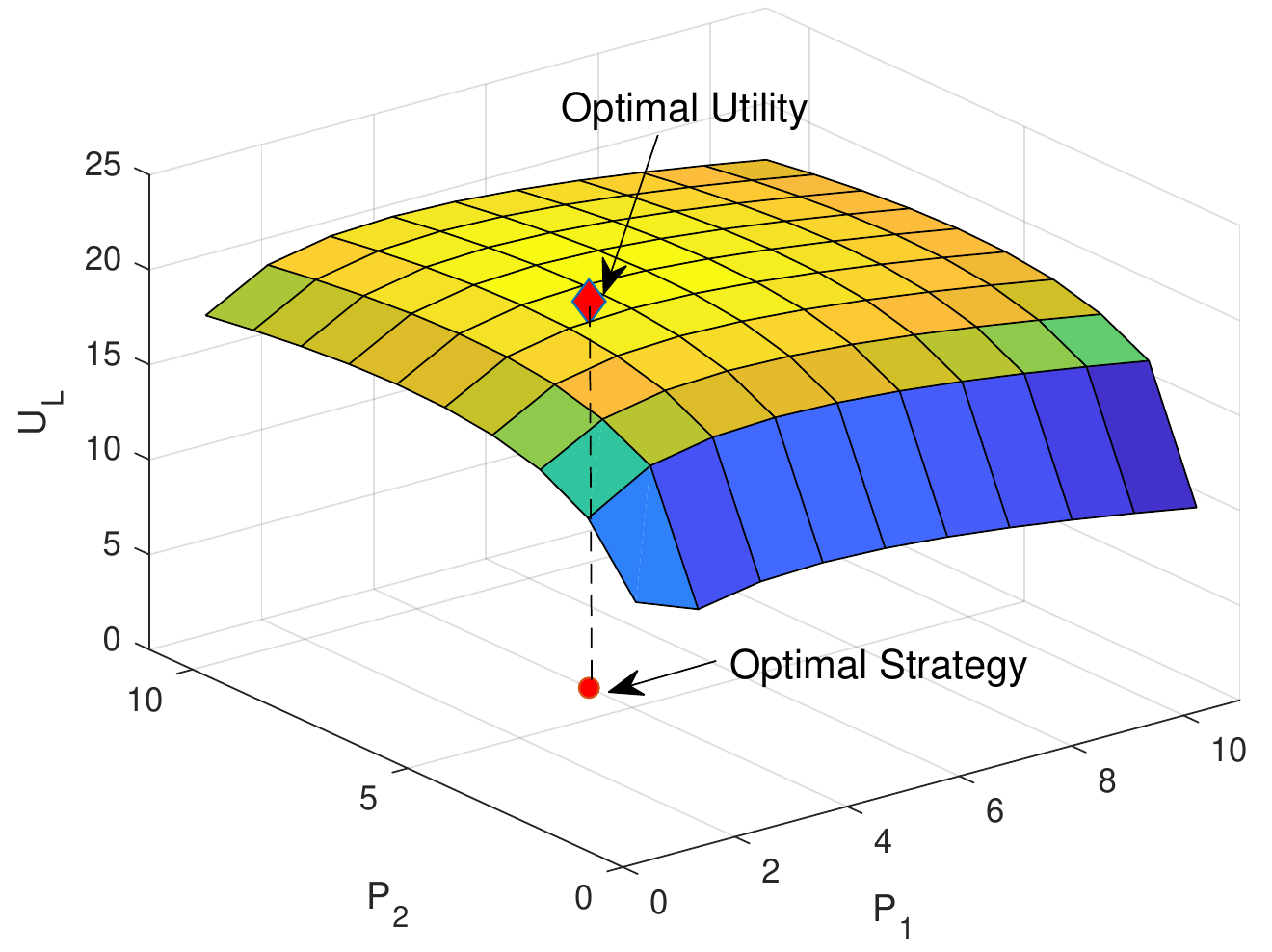}
		\centering
		\caption{MSP's utility function with $\boldsymbol{\alpha}=[4,6]$.}
		\label{Fig:lead}
	\end{figure}

\begin{figure}[!]
	\includegraphics[width=.4\textwidth]{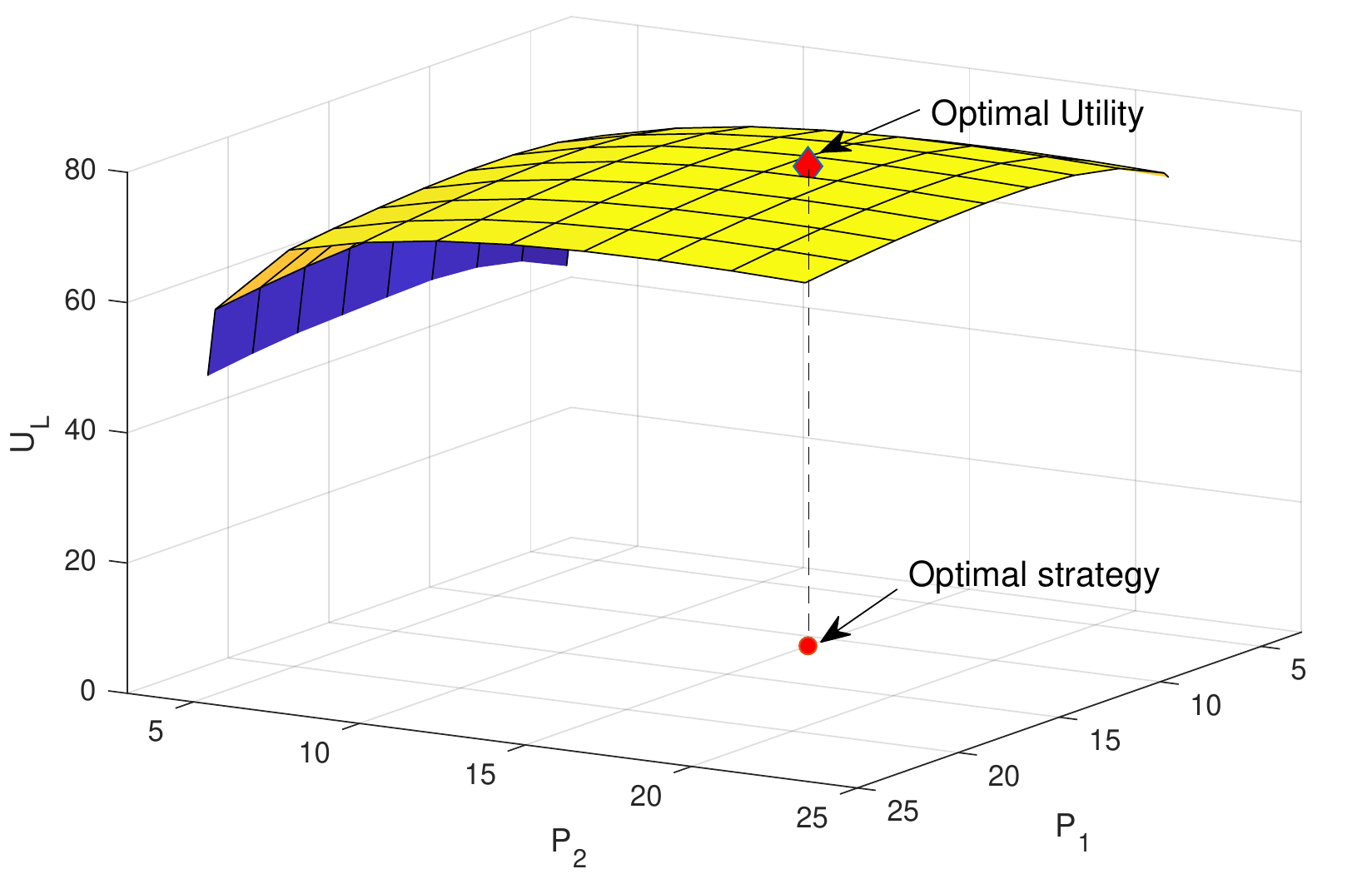}
	\centering
	\caption{MSP's utility function with $\boldsymbol{\alpha}=[10,15]$.}
	\label{Fig:leader2}
\end{figure}

	Fig.~\ref{Fig:lead} illustrates the leader's utility function and optimal strategy. As observed from the figure, the MSP can achieve an optimal utility $U^*_n=20.35$ when it sets $P^*_1=4$ for shard 1 and $P^*_2=6$ for shard 2. It is worth noting that in this example, the MUs do not contribute much resources to the shard. The reason is that the values of $\boldsymbol{\alpha}$ are relatively low in this example, and thus the MSP prioritizes saving costs more. If we increase $\boldsymbol{\alpha}$ to $[10,15]$ as shown in Fig.~\ref{Fig:leader2}, the leader optimal strategy will be increased to $\textbf{P}^*=[11,15]$. This results in an increase of more than 3 times of the resources contributed to the shards. Thus, $\boldsymbol{\alpha}$ is an effective parameter for the MSP to control how much resources it wants at each shard and how much it wants to pay the MUs.
	\section{Conclusion}
	\label{sec:Sum}
	In this paper, we have proposed MetaChain, a novel blockchain-based framework to effectively manage Metaverse applications. MetaChain utilizes the advantages of blockchain and sharding technologies to enable smart and trustful interactions among the MSP and the MUs. Moreover, we have proposed a new sharding scheme that can not only improve the MetaChain performance but also utilize MUs' resources for Metaverse applications. Through Stackelberg game theory analysis, we have examined the MUs' behaviors and developed an economic model that helps the MSP to properly incentivize and allocate MUs' resources based on the Metaverse applications demands. Finally, we have conducted numerical experiments to evaluate the effectiveness of our proposed economic model. For future research direction, the game model can be further extended to a multi-leader-multi-follower Stackelberg game to capture more complex interactions among multiple MSPs and MUs.
	
	
%
	
	\vspace{-0.35em}
	
	\bibliographystyle{IEEE}

\begin{thebibliography}{1}

		\bibitem{meta2} 
	L. H. Lee et al., ``All one needs to know about Metaverse: A complete survey on technological singularity, virtual ecosystem, and research agenda,'', Oct. 2021 \emph{ arXiv preprint arXiv:2110.05352}. [Online]. Available: https://arxiv.org/abs/2110.05352
			\bibitem{web}
	S. Kovach, ``Next for the Metaverse: Convincing you it’s not just for kids,'' \emph{CNBC}, Dec. 22, 2021. [Online]. Available: https://www.cnbc.com/2021/12/22/here-are-the-companies-building-the-Metaverse-meta-roblox-epic.html. [Accessed: 24-Dec-2021].

		\bibitem{re2}
	Y. Han, D. Niyato, C. Leung, C. Miao and D. I. Kim, ``A dynamic resource allocation framework for
	synchronizing Metaverse with IoT service and data,'' Oct. 2021, \emph{arXiv preprint arXiv:2111.00431}. [Online]. Available: https://arxiv.org/abs/2111.00431
		\bibitem{re3}
	Y. Jiang et al., ``Reliable coded distributed computing for
	Metaverse services: Coalition formation and
	incentive mechanism design,'' Nov. 2021, \emph{arXiv preprint arXiv:2111.10548}. [Online]. Available: https://arxiv.org/abs/2111.10548
		\bibitem{re4}
	W. C. Ng et al., ``Unified resource allocation framework for the
	edge intelligence-enabled Metaverse,'' Oct. 2021, \emph{arXiv preprint arXiv:2110.14325}. [Online]. Available: https://arxiv.org/abs/2110.14325
	\bibitem{re5}
	W. C. Ng et al., ``Wireless edge-empowered Metaverse: A
	learning-based incentive mechanism for virtual
	reality,'' Nov. 2021, \emph{arXiv preprint arXiv:2111.03776}. [Online]. Available: https://arxiv.org/abs/2111.03776
		\bibitem{Bitcoin}
		S. Nakamoto, (May 2008). ``Bitcoin: A peer-to-peer electronic cash system''. [Online]. Available: https://bitcoin.org/bitcoin.pdf
		\bibitem{PoS}
		C. T. Nguyen, D. T. Hoang, D. N. Nguyen, D. Niyato, H. T. Nguyen, and E. Dutkiewicz, ``Proof-of-Stake Consensus Mechanisms for Future Blockchain Networks: Fundamentals, Applications and Opportunities,'' in \emph{IEEE Access}, vol. 7, pp. 85727-85745, June 2019.
\bibitem{shard}
G. Yu, X. Wang, K. Yu, W. Ni, J. A. Zhang and R. P. Liu, ``Survey: Sharding in blockchains,'' \emph{IEEE Access}, vol. 8, pp. 14155-14181, Jan. 2020.

		 
	
		
		\bibitem{SC}
		L. Luu, D. Chu, H. Olickel, P. Saxena, and A. Hobor, ``Making smart contracts smarter,'' in \emph{Proc. of the ACM SIGSAC Conference on Computer and Communications Security}, Vienna, Austria, Oct. 2016, pp. 254-269.
		
		\bibitem{Game}
		Z. Han, D. Niyato, W. Saad, T. Başar, and A. Hjørungnes, \emph{Game Theory in Wireless and Communication Networks: Theory, Models, and Applications}. Cambridge, UK: Cambridge University Press, 2012.
		\bibitem{pps}
		M. Rosenfeld ``Analysis of bitcoin pooled mining reward systems,'' Dec.2011, \emph{arXiv preprint arXiv:1112.4980}. [Online]. Available: https://arxiv.org/abs/1112.4980
		\bibitem{Rosen}
		J. B. Rosen, ``Existence and Uniqueness of Equilibrium Points for Concave N-Person Games,'' \emph{Econometrica}, vol. 33, no. 3, pp. 520-534, July 1965.
		\bibitem{cong2}
			C. T. Nguyen, D. T. Hoang, D. N. Nguyen, H. Pham, N. H. Tuong and Dutkiewicz E, ``Blockchain-based Secure Platform for Coalition
		Loyalty Program Management,'' \emph{2021 IEEE Wireless Communications and Networking Conference (WCNC)}, Nanjing, China, Mar. 29 - Apr.1 , 2021, pp. 1-6.
	\end{thebibliography}
	
	\appendices
	\section{Proof of Theorem 1}
	Due to the limited space, we explain briefly here how to prove this Theorem. According to~\cite{Game}, if $S_n$ is compact and convex and $u_n$ is quasi-concave $\forall n \in \mathcal{N}$, there exists at least one Nash equilibrium. It is straightforward to see $S_n$ is compact and convex. Then, we can take the second-order derivative of $U_n^m$, which is always negative, and thus $U_n$ is strictly concave. 
	
	Next, we prove the uniqueness of the equilibrium using Rosen's theorem~\cite{Rosen}. According to~\cite{Rosen}, we need to prove $[\textit{G}(\mathbf{s},\omega)+\textit{G}^{T}(\mathbf{s},\omega)]$ is negative definite for a fixed $\omega > 0$.  Similar to the proof of Theorem 4 in~\cite{cong2}, we can rewrite $\textit{G}(\mathbf{s},\omega)$ and $\textit{G}^T(\mathbf{s},\omega)$ as the sums of a negative semi-definite matrix and a negative definite matrix. Therefore, $[\textit{G}(\mathbf{s},\omega)+\textit{G}^{T}(\mathbf{s},\omega)]$ is negative definite, and thus the equilibrium is unique.

\end{document}